\documentclass[fleqn,usenatbib]{mnras}

\usepackage{newtxtext,newtxmath}

\usepackage[T1]{fontenc}
\DeclareRobustCommand{\VAN}[3]{#2}
\let\VANthebibliography\thebibliography
\def\thebibliography{\DeclareRobustCommand{\VAN}[3]{##3}\VANthebibliography}

\newcommand{\teff}{$T_{\rm eff}$}
\newcommand{\logg}{$\log g$}

\newcommand{\kepler}{{\it Kepler}}
\newcommand{\kms}{km\,s$^{-1}$}

\newcommand{\ha}{H$\alpha$}

\usepackage{graphicx}	
\usepackage{amsmath}	
 
\usepackage{amssymb}	

\title[Asteroseismology of RRab variable star EZ Cnc]{Asteroseismology of RRab variable star EZ Cnc from {\sl K}2 photometry and LAMOST spectroscopy}

\author[Jiangtao Wang et al.]{
Jiangtao Wang,$^{1}$
Jian-Ning Fu,$^{1}$\thanks{E-mail: jnfu@bnu.edu.cn}
Weikai Zong,$^{1}$\thanks{E-mail: weikai.zong@bnu.edu.cn}
Jiaxin Wang,$^{1}$
and Bo Zhang$^{1}$
\\
$^{1}$Department of Astronomy, Beijing Normal University, 19 Avenue Xinjiekouwai, Beijing 100875, People's Republic of China
}

\date{Accepted XXX. Received YYY; in original form ZZZ}

\pubyear{2015}

\begin{document}
\label{firstpage}
\pagerange{\pageref{firstpage}--\pageref{lastpage}}
\maketitle

\begin{abstract}
EZ Cnc, or EPIC 212182292, is a non-Blazhko RRab variable star located in the field of K2 Campaign 16.
Its atmospheric parameters (\teff, \logg, [M/H]) and radial velocities are measured from the 55 high-quality LAMOST medium-resolution spectra.
The fundamental frequency  of pulsation is derived as $f=1.8323(17)$~d$^{-1}$ from the K2 light curves. The amplitude ratios $R_{21} = 0.5115(15), 0.490(8)$, $R_{31} = 0.3249(20), 0.279(7)$ and Fourier phase differences $\varphi_{21}=2.7550(20), 2.764(16)$, $\varphi_{31}=5.7194(25), 5.719(31)$ are determined from the Fourier decomposition of K2 light curve and LAMOST radial velocity curve, respectively.
Through the constraints of the parameters, four optimal models are obtained in a time-dependent turbulent convection model survey for EPIC 212182292.
The parameters of EPIC~212182292 are derived as $M=0.48\pm0.03$\,M$_{\odot}$, $L = 42\pm2$\,L$_{\odot}$, \teff $=6846\pm50$\,K , \logg $=2.79\pm0.01$\,dex, and $Z = 0.006\pm0.002$, respectively.
The precisely determined parameters for RRab variable stars like EPIC~212180092 might help to better understand the period-luminosity relationship of RR Lyrae stars.
\end{abstract}




\section{Introduction}
The RR Lyrae (RRL) stars are large-amplitude radial pulsation variable stars, locating at the intersection between the pulsation instability strip and the horizontal-branch on the Hertzsprung-Russell (HR) diagram.
Current research shows that RRL stars evolved from the main sequence of low-mass stars, with helium burning in the core and hydrogen burning in the shell \citep[see, e.g.,][]{2010A&A...519A..64K}.
They have typical masses in the range $0.55-0.80$\,M$_{\odot}$,
with amplitude of apparent magnitude of 0.3\,-\,2.0 mag and periods of 0.2\,-\,1.0 days. 
According to the pulsation modes, RRL stars are divided into RRab (fundamental mode), RRc (first-overtone mode) and RRd (fundamental and first-overtone mode coexisting) \citep{1902AnHar..38....1B}.
Light curves of some RRL stars show amplitude or phase modulations on long timescales, so-called the Blazhko effect \citep{1907AN....175..325B}. Since its discovery in 1907, there has been no commonly accepted theory to explain this phenomenon yet. The current popular theories account for the Blazhko effect including the magnetic oblique rotation models \citep{1993PASJ...45..617S} and the radial resonant rotation models \citep{2001AcA....51....5N}.

Radial pulsations in RRL stars allow one to calculate the hydrodynamic processes with theoretical models.
The advantage is that it can be directly compared with observations, such as light curves and radial velocity (RV) curves \citep{2017EPJWC.15206001M}.
This method was first applied to the RRc pulsator U~Comes and produced stellar parameters consistent with empirical ones \citep{2000ApJ...532L.129B}.
Through nonlinear convective pulsation model by fitting photometric variations,  \cite{2005AJ....129.2257M} derived the stellar parameters for 14 RRL stars in the Large Magellanic Cloud (LMC) which is used to estimate the average distance of LMC.
\cite{2007A&A...474..557M} successfully constructed nonlinear pulsation models to fit the light curves of six RRL stars in the Galactic globular cluster M3, thereby obtaining an acceptable range of stellar parameters and distances.

The hydrodynamic model of radial pulsation has a variety of processing methods describing the energy transfer and the interaction between pulsation and convection.
\cite{1986A&A...160..116K} derived a new formula in the time-dependent turbulent convection model, and used diffusion approximation consistently throughout the model. 
\cite{2008AcA....58..193S} developed the Kuhfuss model and applied it in the stellar pulsation.
\cite{2019ApJS..243...10P} added the code of Smolec to the Radial Stellar pulsation (RSP) of the Modules for Experiments in Stellar Astrophysics \citep[MESA,][]{2011ApJS..192....3P, 2013ApJS..208....4P, 2015ApJS..220...15P, 2018ApJS..234...34P, 2019ApJS..243...10P}.
However, to determine the precise stellar parameters of RRL stars through the time-dependent convection models, reliable stellar parameter ranges and high cadence photometry and spectroscopy are required.
The previous research used stellar parameters derived from empirical formulae to model RRL stars, and constrained the models only based on light curves \citep{2005AJ....129.2257M, 2007A&A...474..557M}

The LAMOST-\kepler\ project \citep{2015ApJS..220...19D, 2018ApJS..238...30Z, 2020RAA....20..167F} and LAMOST-K2 project \citep{2020ApJS..251...27W} were proposed to use the Large Sky Area Multi-Object Fiber Spectroscopic Telescope \citep[LAMOST,][]{2015RAA....15.1095L} to perform spectroscopic follow-up observations for the \textit{Kepler} mission \citep{2010Sci...327..977B} and the K2 mission \citep{2014PASP..126..398H} of the \textit{Kepler} spacecraft.
The second phase of the LAMOST-\kepler\ project has performed time-domain spectroscopic observations for stars in several fields of the \textit{Kepler} and K2 mission during the LAMOST medium-resolution survey \citep{2020ApJS..251...15Z}.
The continuous light curves with high-precision photometery of \textit{Kepler}/K2, stellar atmospheric parameters provided by the LAMOST-\textit{Kepler} and -K2 projects, and the absolute calibration distance of Gaia \citep{2016A&A...595A...1G} allow one to determine the stellar parameters and obtain internal information of variable stars through asteroseismology.
Therefore, asteroseismology of RRL stars based on both the \kepler\ photometry and LAMOST spectroscopic observations may shed light on the study of RRL stars.

EPIC 212182292 (or EZ Cnc, $\alpha_{2000}$ = 08:52:57.70, $\delta_{2000}$ = +23:47:54.20) is a RRab located in the field of K2 Campaign 16. 
The long cadence ($\sim$ 30 min) photometry of approximately 80 days was obtained by the \textit{Kepler} spacecraft from December 7 of 2017 to February 25 of 2018. 
On the other hand, we find 55 high-quality spectra of this star in LAMOST data release 7 and 8 (DR7 and DR8\footnote{\url{http://www.lamost.org}}).
In Section \ref{sec:Obsdata}, we provide the RV curves and stellar atmospheric parameters from the LAMOST spectrum and perform Fourier decomposition on the light curve and RV curves to obtain the characteristic parameters.
The time-dependent turbulent convection for target modeling is introduced in Section \ref{subsec:tb}. 
In Section \ref{subsec:pae}, the range of parameters needed to build the model is estimated 
on the  basis of observations. 
The final results of the 4 optimal models are presented in Section \ref{subsec:res} with discussion given in Section \ref{sec:dis}. A brief summary is finally presented in Section \ref{sec:sum}.

\section{data analysis}
\label{sec:Obsdata}

\subsection{Fourier decomposition of light curves} 
\label{subsec:lc}
\begin{figure}
	\includegraphics[width=\columnwidth]{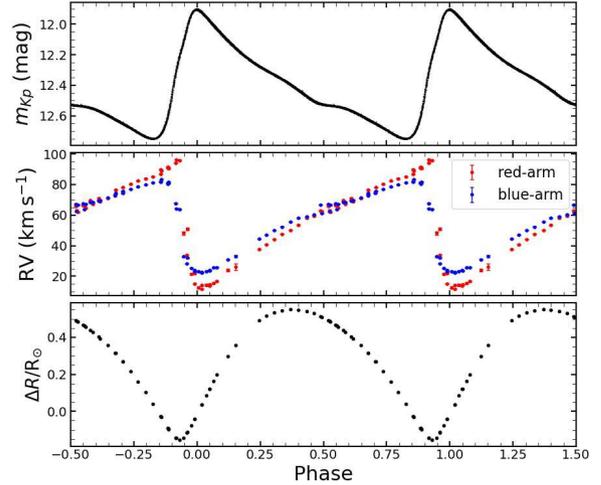}
    \caption{The phase-folded light curve (top panel) and RV curves (middle panel) of EPIC 212182292. The \textit{Kepler} magnitude ($m_{Kp}$) is converted from the {\sl K2} flux with Equation 1.
     The blue and red points in the middle panel represent the RVs measured from the blue-band and the red-band of LAMOST medium-resolution spectra, respectively.
     The bottom panel is the radius amplitude calculated by combining $RV_b$ and Equation 4.}
    \label{fig:lc-rv}
\end{figure}

We use the light curve processed by the EPIC Variability Extraction and Removal for Exoplanet Science Targets pipeline \citep[EVEREST, ][]{2016AJ....152..100L} to convert the Flux to \textit{Kepler} magnitude ($K$p) by the formula \citep{2011MNRAS.417.1022N},
\begin{equation}
m_{K\rm{p}} = m_0 - 2.5\log(\rm{Flux})
\end{equation}
Where the magnitude zero-point $m_0$ = 25.4 is derived 
as the difference between
the mean of $K$p and the instrumental magnitude.
We perform Fourier analysis for the light curves with the code Period04 \citep{2005CoAst.146...53L}, and resolve the fundamental frequency of $f = 1.8323(17)$ day$^{-1}$, which corresponds to the period of $P = 0.5458(5)$ days.
Then the light curves of the EPIC~212182292 folded through the $P$ is shown in the top panel of Figure \ref{fig:lc-rv} where the morphology of the light curve confirms no Blazhko modulation happened.
We note the features of the RRab light curves \citep{2020A&A...635A..66P}, which are a significant bump at $\varphi = 0.50$ and a slight hump at $\varphi = 0.95$.
Note that the zero phases ($\varphi = 0$) is defined at the moment when the light curve maximum bright for the first time.

As \cite{1981ApJ...248..291S} used the Fourier decomposition method to describe the light curve feature of Cepheid variable stars, which was later applied to RRL stars \citep{2011MNRAS.417.1022N}, 
we use the following formula to perform Fourier analysis on the light curves,
\begin{equation}
m(t) = A_0 + \sum_k A_k \sin(2\pi kf(t - t_0) + \varphi_k)
\end{equation}
where $A_0$ is a constant term, $k\in N (k > 0)$.
The $m(t)$ is the apparent magnitude, $t$ the observation time (Barycentric Julian Date: BJD-2454833.0), and $t_0$ the time of the first maximum apparent magnitude of the light curves. 
$A_k$ and $\varphi_k$ represent the amplitude and phase of the $k$th Fourier term, respectively. 
Then, one can calculate the following terms to describe the feature of the light curve.
\begin{eqnarray}
\begin{array}{l}
R_{k1} = A_k/A_1,\\
\varphi_{k1} = \varphi_k - k \varphi_1
\end{array}
\end{eqnarray}
where $k = $ 2 and 3 for the fundamental pulsation of RRL stars \citep{2013MNRAS.428.3034S}.
We calculate the amplitude of fundamental frequency ($A_1$), and the peak-to-peak amplitude ($A$) and rise time ($RT$) from the minimum and the maximum brightness of the phased light curves.

In line to the above method, we, again, have extracted the harmonics up to the third order of the fundamental frequency based on the standard Fourier transform and nonlinear least square fittings. Table \ref{tab:fourier} lists the derived values of those characteristic signals from the {\sl K}2 light curves (Column 2). As a comparison, Table\,\ref{tab:fourier} (Column~3) also list those values estimated from the photometry of the Wide-Angle Search for Planets \citep{2014MNRAS.445.1584S}. Our results are similar but with a higher precision to theirs in terms of the phase. As the different wavelength coverage of the two bands, the quantity amplitude $A$ and $A_1$ are clearly different with each other.

\begin{table}
	\centering
	\caption{The characteristic parameters of the light curves (LC) and radial velocity curves (RVC).}
	\label{tab:fourier}
	\begin{tabular}{llll} 
		\hline
		 & K2 LC & ref.$^b$ & RVC\\
		\hline
		$A$ (mag/km\,s$^{-1}$) & 0.8455(1) & 0.99(18) & 60.83(1.71)  \\
		$A_1$ (mag/km\,s$^{-1}$)& 0.2968(1) & 0.3496(10) & 23.11(24) \\
		$RT$ (rad) & 0.1700(29) & 0.18 & 0.16(2) \\
		$R_{21}$ & 0.5115(15) & 0.5037(32) & 0.490(8)\\
		$R_{31}$ & 0.3249(20) & 0.3170(29) & 0.279(7)\\
		$\varphi _{21}$ (rad) & 2.7550(20) &2.6580(80)& 2.764(16) \\
		$\varphi_{31}$ (rad)& 5.7194(25) & 5.5560(30) & 5.719(32) \\
		\hline
		\multicolumn{4}{l}{\footnotesize$^a$ The parameters are from  \cite{2014MNRAS.445.1584S}}\\
	\end{tabular}
\end{table}

\subsection{Spectroscopic analysis}
\label{subsec:spec}

We obtained 55 high-quality medium-resolution spectra with a signal-to-noise ratio greater than 10 from LAMOST DR7 and DR8. The wavelength coverage of these spectra is divided into the red-arm (the wavelength $\lambda \in [630, 680]$\,nm) and the blue-arm ($\lambda \in [495, 535]$\,nm).

The radial velocities (RVs) of the red- and blue-arm are derived through the SLAM pipeline \citep{2020RAA....20...51Z}.
Howere, the RVs measurements provided by the LAMOST medium-resolution survey are found to have systematic errors between different spectrographs and exposures \citep{2019RAA....19...75L,2020ApJS..251...15Z}.
Therefore, the RVs are calibrated by following the method of \cite{2021arXiv210511624Z}.

The RV curves for EPIC 212182292 obtained from the medium-resolution spectra are shown in the middle panel of Figure \ref{fig:lc-rv},
Where the red dots represent the RVs measured from the \ha\ lines of the red-arm spectra ($\rm{RV_{r}}$, $\lambda \in [630, 680]$\,nm), and the blue dots represent the radial velocities (RVs) measured from the metal lines of the blue-arm spectra ($\rm{RV_{b}}$).
Table \ref{tab:rv} lists the $\rm{RV_{r}}$ and $\rm{RV_{b}}$, from which one finds significant differences between the $\rm{RV_{r}}$ and $\rm{RV_{b}}$. 
The characteristic parameters of the RV curve provided by the blue-arm spectra are listed in the third column of Table \ref{tab:fourier}.
The peak-to-peak amplitude of the $\rm{RV_{b}}$ curve is smaller than that of the $\rm{RV_{r}}$ curve by 23.77\,\kms.
At the phase of 0.90, $\rm{RV_{r}}$ and $\rm{RV_{b}}$ have the largest difference of 32.05\,\kms.

\begin{table}
	\centering
	\caption{The radial velocities of EZ Cnc measured from the red-arm spectra and the blue-arm spectra of LAMOST medium-resolution survey.}
	\label{tab:rv}
	\begin{tabular}{llll} 
		\hline
		BJD & phase & $\rm{RV_{r}}$ & $\rm{RV_{b}}$  \\
		(day)   &  &  (\kms)       & (\kms)    \\
		\hline
        2458538.1608  & 0.0059  & 12.68$\pm$0.48  & 23.08$\pm$0.13 \\
        2458567.0938  & 0.0190  & 11.72$\pm$0.26  & 22.64$\pm$0.07 \\
        2458919.1156  & 0.0192  & 14.32$\pm$0.47  & 22.21$\pm$0.13 \\
        2458538.1775  & 0.0365  & 14.36$\pm$0.49  & 23.36$\pm$0.12 \\
        2458567.1104  & 0.0495  & 13.74$\pm$0.27  & 23.95$\pm$0.07 \\
        2458919.1322  & 0.0498  & 14.43$\pm$0.39  & 23.51$\pm$0.09 \\
        2458538.1935  & 0.0657  & 15.62$\pm$0.47  & 25.61$\pm$0.13 \\
        2458919.1482  & 0.0790  & 16.86$\pm$0.45  & 25.83$\pm$0.13 \\
        2458502.2036  & 0.1224  & 24.13$\pm$0.78  & 31.03$\pm$0.24 \\
        2458502.2196  & 0.1517  & 26.24$\pm$2.12  & 33.08$\pm$0.55 \\
        2458536.1081  & 0.2448  & 37.55$\pm$0.23  & 44.48$\pm$0.07 \\
        2458536.1241  & 0.2741  & 40.82$\pm$0.24  & 47.09$\pm$0.08 \\
        2458536.1408  & 0.3046  & 44.12$\pm$0.21  & 50.00$\pm$0.08 \\
        2458536.1574  & 0.3352  & 47.71$\pm$0.18  & 52.36$\pm$0.07 \\
        2458536.1769  & 0.3708  & 50.24$\pm$0.29  & 55.54$\pm$0.09 \\
        2458536.1929  & 0.4001  & 53.47$\pm$0.25  & 58.01$\pm$0.07 \\
        2458448.3420  & 0.4333  & 58.00$\pm$0.45  & 59.00$\pm$0.15 \\
        2458448.3587  & 0.4639  & 59.55$\pm$0.37  & 60.30$\pm$0.19 \\
        2458558.0722  & 0.4890  & 63.02$\pm$0.36  & 66.73$\pm$0.15 \\
        2458448.3747  & 0.4931  & 62.37$\pm$0.56  & 61.19$\pm$0.19 \\
        2458558.0889  & 0.5196  & 65.93$\pm$0.37  & 67.09$\pm$0.11 \\
        2458448.3914  & 0.5237  & 62.98$\pm$0.64  & 62.33$\pm$0.24 \\
        2458540.0830  & 0.5278  & 62.09$\pm$0.31  & 66.55$\pm$0.09 \\
        2458558.1049  & 0.5488  & 66.48$\pm$0.32  & 67.61$\pm$0.15 \\
        2458448.4073  & 0.5529  & 66.17$\pm$0.33  & 63.80$\pm$0.26 \\
        2458540.0996  & 0.5583  & 64.66$\pm$0.20  & 67.36$\pm$0.07 \\
        2458558.1208  & 0.5781  & 69.66$\pm$0.26  & 68.54$\pm$0.13 \\
        2458540.1156  & 0.5876  & 67.28$\pm$0.30  & 68.35$\pm$0.08 \\
        2458558.1382  & 0.6099  & 70.79$\pm$0.50  & 69.40$\pm$0.17 \\
        2458540.1316  & 0.6169  & 69.06$\pm$0.26  & 69.81$\pm$0.12 \\
        2458540.1482  & 0.6474  & 71.12$\pm$0.29  & 71.23$\pm$0.09 \\
        2458540.1642  & 0.6767  & 73.24$\pm$0.31  & 72.82$\pm$0.11 \\
        2458539.0733  & 0.6778  & 76.47$\pm$0.22  & 74.20$\pm$0.06 \\
        2458540.1802  & 0.7059  & 75.09$\pm$0.37  & 74.28$\pm$0.10 \\
        2458539.0893  & 0.7071  & 78.09$\pm$0.21  & 75.66$\pm$0.07 \\
        2458539.1052  & 0.7363  & 80.27$\pm$0.20  & 77.09$\pm$0.08 \\
        2458539.1219  & 0.7669  & 82.32$\pm$0.24  & 78.56$\pm$0.09 \\
        2458539.1379  & 0.7961  & 83.90$\pm$0.34  & 80.02$\pm$0.11 \\
        2458539.1538  & 0.8254  & 85.27$\pm$0.34  & 81.14$\pm$0.10 \\
        2458539.1705  & 0.8559  & 89.94$\pm$0.25  & 81.90$\pm$0.12 \\
        2458567.0049  & 0.8561  & 86.87$\pm$0.27  & 81.64$\pm$0.09 \\
        2458538.0803  & 0.8583  & 89.10$\pm$0.25  & 83.44$\pm$0.11 \\
        2458920.1336  & 0.8845  & 90.94$\pm$0.47  & 81.08$\pm$0.21 \\
        2458539.1865  & 0.8852  & 91.41$\pm$0.39  & 80.38$\pm$0.17 \\
        2458567.0215  & 0.8867  & 90.66$\pm$0.25  & 80.87$\pm$0.08 \\
        2458538.0963  & 0.8876  & 90.66$\pm$0.31  & 81.73$\pm$0.14 \\
        2458567.0375  & 0.9159  & 94.25$\pm$0.25  & 67.70$\pm$0.14 \\
        2458538.1122  & 0.9169  & 96.30$\pm$0.27  & 64.25$\pm$0.20 \\
        2458919.0670  & 0.9302  & 95.61$\pm$0.42  & 63.92$\pm$0.21 \\
        2458538.1289  & 0.9474  & 48.23$\pm$1.12  & 32.88$\pm$0.12 \\
        2458567.0618  & 0.9604  & 33.86$\pm$0.40  & 28.18$\pm$0.09 \\
        2458919.0836  & 0.9607  & 50.91$\pm$1.05  & 32.02$\pm$0.12 \\
        2458538.1449  & 0.9767  & 21.56$\pm$0.41  & 25.35$\pm$0.14 \\
        2458567.0778  & 0.9897  & 15.21$\pm$0.24  & 22.98$\pm$0.06 \\
        2458919.0996  & 0.9900  & 21.80$\pm$0.51  & 24.12$\pm$0.10 \\
		\hline
	\end{tabular}
\end{table}

Since the stellar parameters and spectral types of {RRL stars} vary with the pulsation cycle, it is not easy to obtain reliable parameters from the spectrum \citep{2014MNRAS.445.4094F}.
\cite{2010A&A...519A..64K} claimed that the maximum radius is the best phase for performing 'classical' (i.e. assuming LTE, plane-parallel geometry and a static model atmosphere) spectral analysis, and the stellar parameters can be accurately determined by the equivalent width method \citep{2014MNRAS.445.4094F}.
The periodic variations of RV contain the information of the radius amplitude of pulsating stars and can be calculated by the following formula \citep{2013MNRAS.434..552C},
\begin{equation}
\Delta R(t) = \int_{0}^{P} p\,(RV(t) - {\rm RV_0})\, dt
\end{equation}
Where $P$ is the period and $\rm{RV_0 = 69.26\ km\,s}^{-1}$ is the RV of the center-of-mass of the star, for which we take the mean of the RV curve \citep{2013MNRAS.434..552C}.
The $p = 1.25$ is a projection factor between the observed RVs and the pulsation velocity of star \citep{2017A&A...597A..73N}.
Based on the observed data, we calculate the radius amplitude, $\Delta R(t)$,  and derive the maximum radius amplitude, $\Delta R(t)_{\rm{max}}= 0.704(6) \,\rm{R_{\odot}}$.
The relationship between the radius amplitude and the pulsation phase is shown in the bottom panel of Figure \ref{fig:lc-rv}.
The radius reaches the maximum at $\varphi = 0.40$.
Therefore, we select 7 blue-arm spectra with phases between 0.30 and 0.50 to determine the stellar atmospheric parameters through 'equivalent widths method' tool of iSpec \citep{2014ASInC..11...85B, 2014A&A...569A.111B}.
We average the measured results to obtain the stellar atmospheric parameters of EPIC 212182292, as \teff\ = 6569$\pm$200\,K, \logg\ = 2.46$\pm$ 0.05\,dex, and [M/H] = -0.40$\pm$0.55\,dex.

\section{Numerical modelling}
\subsection{Theoretical modelling}
\label{subsec:tb} 

\cite{2008AcA....58..193S} implemented the stellar radial pulsation convective code based on the time-dependent turbulent convection model proposed by \cite{1986A&A...160..116K}.
The code couples the convection and the pulsation driven by the partial ionization of H and He, which can effectively reproduce the light curves and the RV curves of classical pulsating variable stars.
The turbulent energy and the kinetic energy are coupled to each other through coupling terms \cite[][equation 11]{2008AcA....58..193S} controlled by the eight order of unity convection parameters.
In the models, different physical mechanisms can be considered by setting the convection parameters, as listed in Table \ref{tab:alpha}, which contains eight following rows \citep{2019ApJS..243...10P}:
\\
(1) $\alpha$: the mixing-length parameter;\\
(2) $\alpha_m$: the eddy-viscous dissipation parameter;\\
(3) $\alpha_s$: the turbulent source parameter;\\
(4) $\alpha_c$: the convective flux parameter ;\\
(5) $\alpha_d$: the turbulent dissipation parameter;\\
(6) $\alpha_p$: the turbulent pressure parameter;\\
(6) $\alpha_t$: the turbulent flux parameter;\\
(8) $\gamma_r$: the radiative cooling parameter.\\
Most values of these convection parameter sets of C1, C2, C3 and C4 are recommended to be set as default by \cite{2019ApJS..243...10P} except the $\alpha_m$ to be modified a bit in C3 and C4. The description of the different physical treatments for those four sets of models can be found in Section~2.2.5 of \citet{2019ApJS..243...10P}.

The Smolec code is integrated in the Radial Stellar Pulsations (RSP) module of MESA environment \citep{2011ApJS..192....3P,2013ApJS..208....4P,2015ApJS..220...15P,2018ApJS..234...34P,2019ApJS..243...10P} and used to model classic large-amplitude, pulsation variable stars. 
The RSP module first builds an initial model concerning the stellar parameters (mass ($M$), effective temperature (\teff), luminosity ($L$), hydrogen abundance ($X$), and metallicity ($Z$)). It is a chemically homogeneous  envelope structure and independent of the detailed core structure.
Then, it performs linear non-adiabatic stability analysis on the initial model to derive the eigenmodes with periods.
The initial model is then perturbed by eigenvectors, and the time-dependent nonlinear equations \citep[][equation 1--4]{2019ApJS..243...10P}, finally are integrated for time evolution calculations.
\begin{table} 
	\centering
	\caption{The C1, C2, C3 and C4 convection parameter sets taken in the models.}
	\label{tab:alpha}
	\begin{tabular}{lllll} 
		\hline
		 parameter & C1 & C2 & C3 & C4 \\
		\hline
		$\alpha$ & 1.5 & 1.5 & 1.5 & 1.5 \\
		$\alpha_m$ & 0.25 & 0.50 & 0.25 & 0.50 \\
		$\alpha_s$ & 1.0 & 1.0 & 1.0 & 1.0 \\
		$\alpha_c$ & 1.0 & 1.0 & 1.0 & 1.0 \\
		$\alpha_d$ & 1.0 & 1.0 & 1.0 & 1.0 \\
		$\alpha_p$ & 0.0 & 0.0 & 1.0 & 1.0 \\
		$\alpha_t$ & 0.00 & 0.00 & 0.01 & 0.01 \\
		$\gamma_r$ & 0.0 & 1.0 & 0.0 & 1.0 \\
		\hline
	\end{tabular}
\end{table}

\subsection{Parameter estimation}
\label{subsec:pae}
The calibrated Gaia DR2 distance of EPIC 212182292 is $r = 1840_{-161}^{+192}$\,pc \citep{2018AJ....156...58B}. We calculate the bolometric luminosity ($L$) by the formula \citep{2018AJ....156...58B},
\begin{eqnarray}
\begin{array}{l}
\label{eq:lum}
M_{g} = m_{g} + 5(1 - \log{r}) - A_{g},\\
-2.5\log{L} = M_{Kp} + BC_{g}(T_{\rm eff}) - {\rm M_{bol,\odot}}
\end{array}
\end{eqnarray}
Where $M_g$ and $m_g = 12.2989$ are the absolute magnitude and apparent magnitude in the  Gaia G band, respectively. 
As the absolute bolometric magnitude of the Sun $\rm{M_{bol_{\odot}}} = 4.74$ is defined by IAU \citep{2015arXiv151006262M}, $A_{g} = 0.1069$ is the extinctions in G band \citep{1989ApJ...345..245C, 1994ApJ...422..158O},  
$BC_{g}$(\teff) is the temperature-only dependent bolometric correction \citep{2018A&A...616A...8A}. 
we calculate the bolometric luminosity of star by substituting the observation parameters into Formula \ref{eq:lum} to get $L= 34.5\pm 7.5\,{\rm L_{\odot}}$.

The metallicity is calculated by the approximation \citep{2012MNRAS.427..127B},
\begin{eqnarray}
\begin{array}{l}
{\rm [M/H]}=\log{(Z/X)}-\log{(Z/X)_{\odot}},\\
Y=0.2485+1.78Z,\\
X + Y + Z = 1
\end{array}
\end{eqnarray}
Where $\rm{(Z/X)_{\sun}=0.0207}$ and Y is the helium abundance. 
We calculate the hydrogen abundance $X = 0.735(24)$, and the metallicity $Z = 0.006(9)$.
We use the mass of EPIC 212182292, $M = 0.51\pm0.02$ M$_\odot$, provided in \cite{2014MNRAS.445.1584S} as an initial reference.
In Table \ref{tab:aps} we list the estimated stellar parameters.


\begin{table} 
	\centering
	\caption{The estimated stellar parameters of EPIC 212182292}
	\label{tab:aps}
	\begin{tabular}{ll} 
		\hline
		Parameter & Value\\
		\hline
		\teff (K)            & 6569(200)\\
		\logg (dex)          & 2.46(5)\\
		X                    & 0.735(24)\\
		Z                    & 0.006(9)\\
		$L (L$\sun$)$        & 34.5(7.5)\\
		$M$ (M$\sun$)        & 0.51(2)$^a$\\
		\hline
		\multicolumn{2}{l}{\footnotesize$^a$ \cite{2014MNRAS.445.1584S}}\\
	\end{tabular}
\end{table}

\subsection{Model searching}
\label{subsec:res}
 

We use the MESA-RSP module \citep[version 15140;][]{2019ApJS..243...10P} to search for models for EPIC~212182292.
The constructed models adopt fundamental mode pulsations and the amplitude of the initial surface velocity perturbation is 4.5\,\kms.
We consider the convection parameter set C1 to construct the model grid based on the stellar parameters estimated in section
\ref{subsec:pae}.
The stellar parameters of the grid are as follows,\\
$M$: [0.40, 0.65]\,M$_{\odot}$ with the step of 0.01\,M$_{\odot}$,\\
\teff: [6200, 7950]\,K with the step of 50\,K,\\
$L$: [25, 55]\,L$_{\odot}$ with the step of 1\,L$_{\odot}$,\\
$[{\rm M/H}]$: [-1.1, 0]\,dex , with the step of 0.1 dex.\\
The coupled $X$ and $Z$ are calculated according to [M/H] and Equation 6.

\begin{figure}
	\includegraphics[width=\columnwidth]{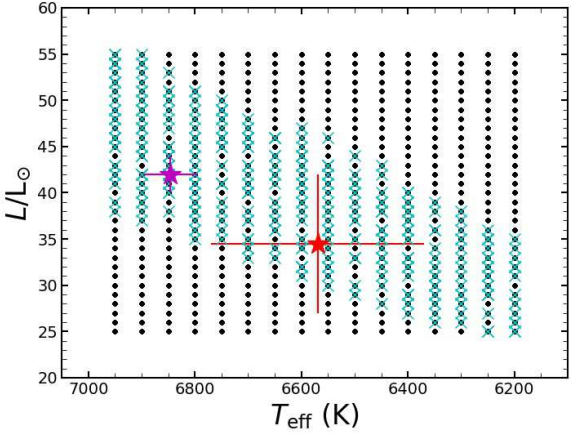}
    \caption{The positions of the model grid in the HR diagram. Black dots represent all models in the grid. Cyan crosses represent the models selected by the period matching. The red asterisk represents EPIC 212182292 from observational determination.
    The magenta asterisk indicates the position of the weighted average results from four optimal models (see Section~\ref{subsec:res} for details).
    }
    \label{fig:HR}
\end{figure}

A total of $154752 = 26 \times 16 \times 31 \times 12$ models in the grid and the fundamental periods are obtained through the linear non-adiabatic stability analysis of MESA-RSP module.
First we use the period constraint to ensure that the period offsets between the models and the observations are smaller than the uncertainty of the observed period $\Delta P = 0.0005$ days.
Figure \ref{fig:HR} shows the positions of the model grid and 356 constrained models (with parameters \teff\ and $L$) in the HR diagram.
One should note that each point denotes a set of models with the above three different parameters other than $L$ and \teff.

As the structures of light curves from the constructed models varied with different convection parameter sets, we build a series of models ($4\times 356$) to calculate the complete light curves and RV curves by taking the C1--C4 convection parameter sets.
In order to have consistency between the model light curves and the K2 light curves, the bolometric light curves of the models are transformed to the \kepler\ band through the bolometric calibration coefficient given by \cite{2019MNRAS.489.1072L}, which depends only on the effective temperature.
We calculate the offsets of the models and the observation measurements in the characteristic parameter space wih the following formula \citep{2013MNRAS.428.3034S}.
\begin{equation}
d = \sqrt{\sum _i (W_{i,\rm{mod}} - W_{i,\rm{obs}})^2/W_{i,\rm{obs}}^2},
\end{equation}
Where $W_{i}$ represents $A$, $A_1$, $RT$, and four Fourier parameters ($R_{21}, R_{31}, \varphi_{21}, \varphi_{31}$) for model curves ( $W_{i,{\rm mod}}$) and observed curves ( $W_{i,{\rm obs}}$, Columns 2 and 4 of Table \ref{tab:fourier}).
The smaller $d$ indicates that the light curve and the RV curve of the model are closer to those of EPIC~212182292.

\begin{figure}
	\includegraphics[width=\columnwidth]{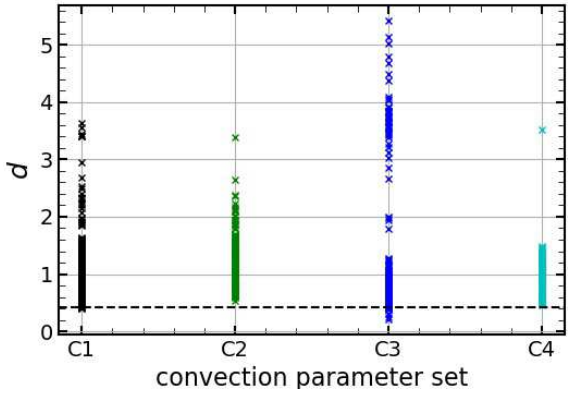}
    \caption{ 
    The $d$ values of different convection parameter sets of C1, C2, C3, and C4.
    Each convection parameter set consist of 356 selected models by the observed period matching.
    The black dashed line is the position of $d = 3d_{\rm err}$ = 0.42.
    }
    \label{fig:cd}
\end{figure}

\begin{figure}
	\includegraphics[width=\columnwidth]{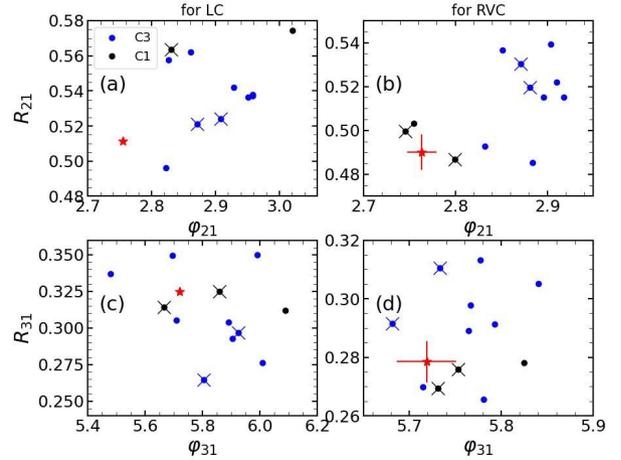}
    \caption{The Fourier parameters R$_{21}$ vs. $\varphi _{21}$ (top panels), and R$_{31}$ vs. $\varphi _{31}$ (bottom panels) plots of the light curves and RV curves from the candidate models and observations, respectively.
    The mark colors indicate the corresponding convection parameter groups in the legend. The cross represents 4 optimal models (M4, M5, M8, and M12), the red asterisk represents observations determined values.}
    \label{fig:rp}
\end{figure}

Figure \ref{fig:cd} shows the values of $d$ for the $4\times356$ models with C1, C2, C3 and C4 convection parameter sets. 
The error of $d$ is $d_{\rm err} = 0.14$, which is estimated from the uncertainties of those characteristic parameters.
12 models constructed with either C1 or C3 convection parameter sets are selected as the candidates concerning $d\leq 3d_{\rm err}$.
The parameters of these models and the calculated $d$ values are listed in columns 3-5 and 7 of Table \ref{tab:bmaps}.
The Fourier parameters of the model light curves are shown in the left panels of Figure \ref{fig:rp}, and those of the RV curves are shown in the right panels.

\begin{figure}
	\includegraphics[width=\columnwidth]{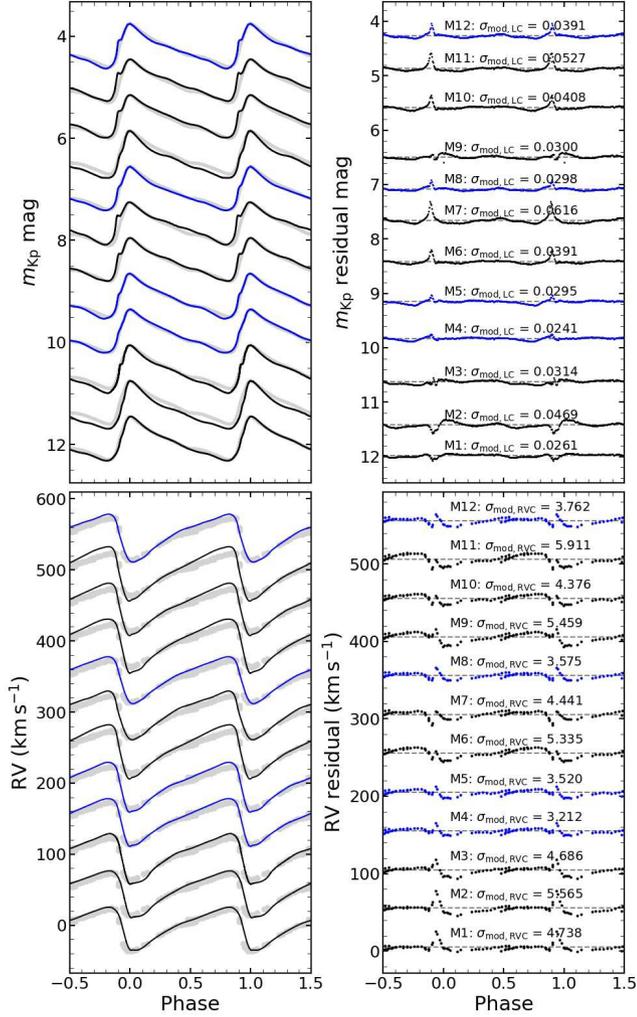}
    \caption{
    Comparison between the observed light curves (top panels) and RV curves (bottom panels) and the ones of 12 models (M1--M12). The light curves and RV curves were shifted by 0.7 mag and 50 \kms for clarity, respectively.
    The light gray dots represent the curves from the observations for EPIC 212182292.
    The blue and black curves represent the curves of the optimal models and other candidate models, respectively.
    The dashed lines in the right panels represent the means of the residuals.}
    \label{fig:lcv}
\end{figure}

Figure \ref{fig:lcv} shows the light curves (top panels) and RV curves (bottom panels) of these models.
The left panels show a comparison between the model curves and the observation curves, and the right panels show the residuals with the corresponding standard deviations $\sigma_{\rm mod, LC}$ for the light curves and $\sigma_{\rm mod, RVC}$ for the RV curves indicated.
In order to select optimal models among these 12 models,
we adopt the standard deviations ($\sigma_{\rm obs}$) between the observation curves and the third-order Fourier fitting curves as the criterion.
These standard deviations are $\sigma_{\rm obs, LC}=0.0406$\,mag for light curves and $\sigma_{\rm obs, \textsc{RVC}}=4.236$\,\kms for RV curves, respectively.
We finally obtained four optimal models by the differences between the observational and model derived results satisfied $\frac{\sigma_{\rm mod, RVC}}{\sigma_{\rm obs, RVC}}\leqslant 1$ and $\frac{\sigma_{\rm mod, LC}}{\sigma_{\rm obs, LC}}\leqslant 1$.
Those optimal models (M4, M5, M8, and M12) are indicated by the blue curves in Figure \ref{fig:lcv}. The contrast suggests that the theoretical curves are somewhat consistent with the observational results.

The properties of EPIC~212182292 are determined through the weighted average parameters of those four models, where the weight $w$ defines as,
\begin{equation}
w = 1/(\sigma_{\rm{mod,RVC}}^2+\sigma_{\rm{mod,LC}}^2),
\end{equation}
which gives $M=0.48\pm0.03$\,M$_{\odot}$, $L = 42\pm2$\,L$_{\odot}$, \teff $=6846\pm50$\,K, and $Z = 0.006\pm0.002$. One can see other parameters from the bottom row of Table \ref{tab:bmaps}.
The location suggests that EPIC~212182292 is hotter and brighter than the determination from the spectroscopic results on the HR~diagram (see Figure \ref{fig:HR} for comparison).

\begin{table*} 
	\centering
	\caption{The physical parameters of the 12 the candidate models for the EPIC 212182292 {\bf ordered by their $d$}.}
	\label{tab:bmaps}
	\begin{tabular}{llllllllllll} 
		\hline
		model & set & $M$ & \teff & $L$ & $X$ & $Z$ & \logg & $\Delta R_{\rm max}$ & $d$ & $\sigma_{\rm mod,LC}$ & $\sigma_{\rm mod,RVC}$ \\
		             &     & (M$_{\odot}$)  & (K)  & (L$_{\odot}$) &  &     & (dex)    & R$_{\odot}$  & & (mag) & (\kms) \\
		\hline
	    M1     &  C3   &  0.45     &  6900  &  42  & 0.748 & 0.002  &  2.771   &  0.638 & 0.206 & 0.0261 & 4.738 \\
	    M2     &  C3   &  0.52     &  6900  &  47  & 0.748 & 0.002  &  2.784   &  0.699 & 0.244 & 0.0469 & 5.565 \\
	    M3     &  C3   &  0.44     &  6850  &  40  & 0.748 & 0.002  &  2.764   &  0.793 & 0.312 & 0.0314 & 4.686 \\
	    M4$^b$     &  C3   &  0.49     &  6800  &  41  & 0.737 & 0.008  &  2.795   &  0.701 & 0.373 & 0.0241 & 3.212 \\
	    M5$^b$     &  C3   &  0.47     &  6800  &  40  & 0.741 & 0.006  &  2.786   &  0.752 & 0.387 & 0.0295 & 3.520 \\
	    M6     &  C3   &  0.62     &  6650  &  44  & 0.727 & 0.014  &  2.826   &  0.787 & 0.393 & 0.0391 & 5.335 \\
	    M7     &  C1   &  0.44     &  6850  &  40  & 0.748 & 0.002  &  2.772   &  0.781 & 0.397 & 0.0616 & 4.441 \\
	    M8$^b$     &  C1   &  0.52     &  6900  &  46  & 0.741 & 0.006  &  2.799   &  0.704 & 0.403 & 0.0298 & 3.575 \\
	    M9     &  C3   &  0.65     &  6750  &  49  & 0.734 & 0.010  &  2.825   &  0.799 & 0.411 & 0.0300 & 5.459 \\
	    M10    &  C3   &  0.52     &  6700  &  40  & 0.734 & 0.010  &  2.804   &  0.761 & 0.411 & 0.0408 & 4.376 \\
	    M11    &  C3   &  0.65     &  6600  &  44  & 0.723 & 0.016  &  2.834   &  0.783 & 0.412 & 0.0527 & 5.911 \\
	    M12$^b$    &  C1   &  0.45     &  6900  &  42  & 0.748 & 0.002  &  2.774   &  0.756 & 0.415 & 0.0391 & 3.762 \\
	    \hline
	    Average$^a$     &    &  0.48$\pm$0.03  &  6846$\pm$50  &  42$\pm$2  & 0.741$\pm$0.004 & 0.006$\pm$0.002  &  2.79$\pm$0.01   &  0.728$\pm$0.026 & & \\
		\hline
		\multicolumn{12}{l}{\footnotesize $^a$ The values are obtained by the weighted average of the models with label $b$.}\\
	\end{tabular}
\end{table*}

\section{Discussion}
\label{sec:dis}
\subsection{\bf Comparison with evolutionary models}
\label{subsec:em}
We evoke that the dynamical model provide a bit lower mass, $0.48\,M_\odot$, than the typical evolutionary mass of RRL stars, $0.55 \sim 0.80 M_\odot$. In order to direct compare to evolutionary models quantitatively, the BaSTI\footnote{A Bag of Stellar Tracks and Isochrones \citep{2004ApJ...612..168P, 2009ApJ...697..275P}, which can be found through the link: basti.oa-teramo.inaf.it} database was used to calculate the properties of 
zero-age horizontal branch (ZAHB) with three different chemical compositions $(Z=0.004,Y=0.251$; $Z=0.008,Y=0.256$; and $Z=0.010,Y=0.259)$. All models evolved through the red giant branch of $0.6{\rm M_{\odot}}$ to ZAHB phase, with input parameters $\alpha = 1.5$ and $\eta = 0.2$, where the latter is the parameter define the mass loss efficiency \citep{1975MSRSL...8..369R}.

\begin{figure}
	\includegraphics[width=\columnwidth]{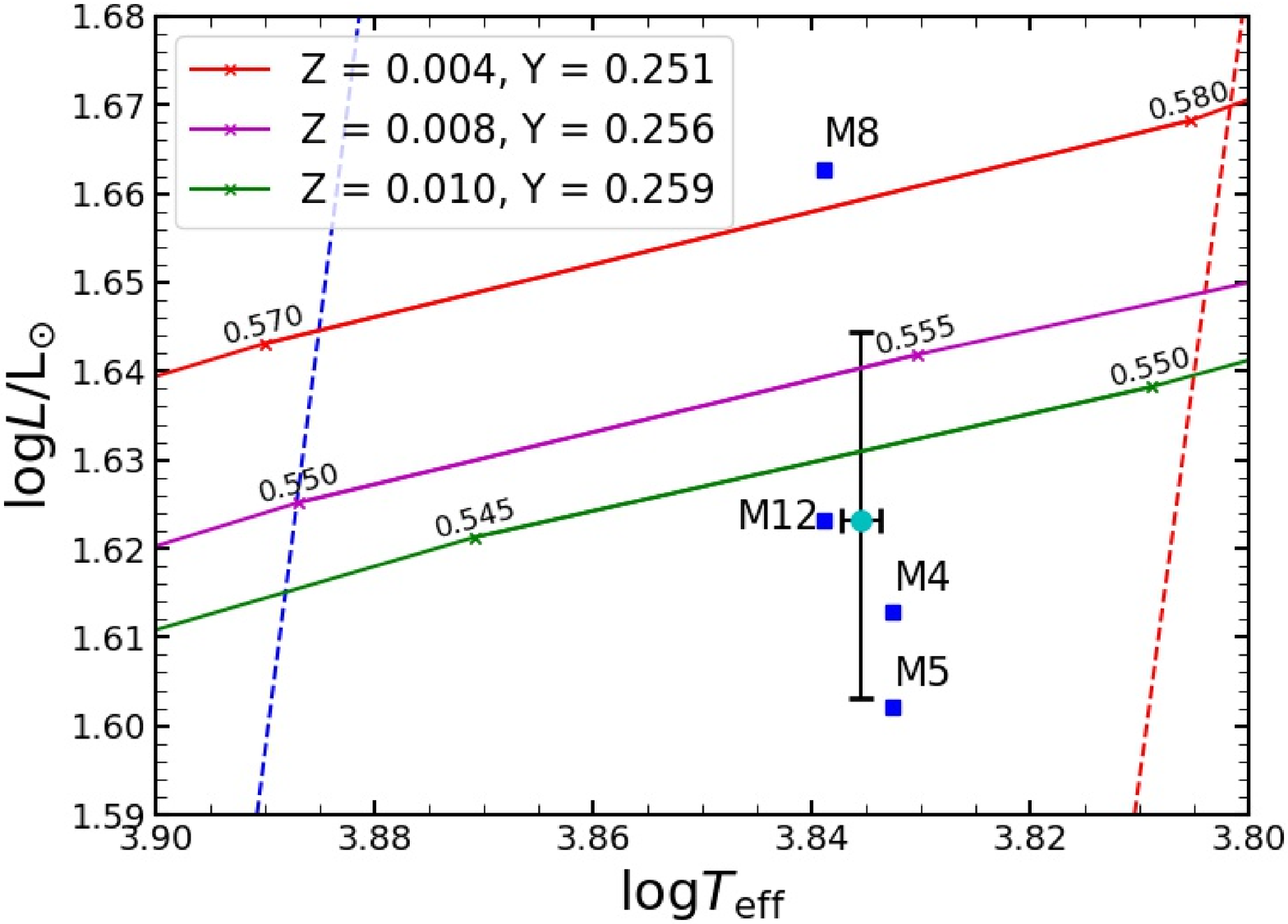}
    \caption{
    Comparison of the location of the optimal models with the ZAHB models on HR~diagrams.
    The solid lines are the isochrone tracks of ZAHB, where the cross marks the mass of the individual model as labeled by text with unit of M$_{\odot}$.
    The square and dot represent the optimal model as labeled by text and the weighted averaged model, respectively. The dashed lines define the blue and red edge of the instability strip \citep{2019AstL...45..353F}.}
    \label{fig:hre}
\end{figure}

Figure\,\ref{fig:hre} presents the contrast of four dynamical models along with the evolutionary results. With the same \teff, the luminosity increases as the metallicity decreases as seen from the ZAHB isochrone tracks. Comparing to the evolutionary results, the models M4, M5, and M12 produce a lower mass and fainter brightness if the chemical abundances and \teff{} are similar. All those three models are found with $M\sim 0.45-0.49~{\rm M_{\odot}}$ and $L\sim 40-42~L\odot$, that are $0.07\sim 0.09$\,M$_{\odot}$ lighter and $0.01\sim 0.03$\,dex fainter than that from the ZAHB models. However, those properties of M8 are comparable to the evolutionary determinations, whose mass is near the lower boundary of the ZAHB star. We note that $Z=0.006$ is not provided in the BaSTI database for direct comparison.

A similar discrepancy of mass and luminosity had been also found between the evolutionary code and pulsation calculations in 19 non-Blazhko RRab stars from {\sl Kepler} photometry \citep{2011MNRAS.417.1022N}. They suggest that this discrepancy might have the same reason as found in Cepheid variables. For instance, \citet{2006ApJ...642..834K} claim that Cepheid stars have experienced enhanced internal mixing near the convective core along the evolution of the main-sequence phase can account for that. However, the low mass of RRL stars cannot develop a convective core sufficiently. 
An observational support for pulsating mass in this discrepancy is that \cite{2010Natur.468..542P} derived an accurate dynamical mass for a classical Cepheid residing a binary system in LMC whose value agrees well with the pulsation solution. Their results indicate that mass loss may significantly increase during pulsation period as a consequence of radial pulsating motion and shock wave in the atmosphere \citep{2008ApJ...677..483K, 2008ApJ...684..569N}. This scenario may enhance the mass loss in RRL variables as well. Besides, \citet{1997ApJ...479..279B} performs theoretical investigation that RRL stars can have much lower mass to 0.36~$M_\odot$ if considering substantial mass-loss during the RGB phase.
In addition, some rare binary systems, with low mass component below the helium ignition $\sim 0.5M_\odot$, could also display RRL-shape light curves when they evolve across the classical instability strip \citep{2017MNRAS.466.2842K}. \citet{2012Natur.484...75P} discovered such a kind system that contains a component with a mass down to 0.26\,M$_{\odot}$ with RRL-like pulsations. However, EPIC~212182292 had never been present any clue to binary, which needs further analysis to search for long-term binary signals residing or not.

\subsection{Models and convection parameters}
\label{subsec:dis}

We select four optimal models from 12 candidates that are met the criterion as mentioned in last section. The averaged weighted values are consistent with the results derived from spectroscopy considering the measured uncertainties. In addition, RRL stars exhibit large amplitude variations, which can systematically affect the \teff{} and luminosity if the spectra were chosen at different phase. Therefore, in some extent, our models show good agreement with the observational determinations. All these four optimal models are found with convective parameters of C1 and C3 sets. It suggests that the radiative cooling $\alpha_\gamma$ is set to zero to reproduce the light curve and RVs of EPIC~212182292. 

In particular, the model M4 with convection parameter set C3 reproduces the bump of the descending branch ($\varphi \sim 0.6$) of the light curve
(see left panel of Figure \ref{fig:lcv}). 
Compared to the other convection parameter sets, the most variants with which we construct the models are the convective parameters described by the turbulent pressure $\alpha_p = 1.0$, the turbulent flux $\alpha_t = 0.01$ and the radiative cooling $\gamma_r = 0$. Thus, the observational profile can indeed be used to constrain those convective parameters in the RRab star EPIC~212182292.

The projection factor is derived as $p=1.11\pm0.02$ from observation RV curves and the model constrained velocities, which
is an important parameter for the estimation of the distance of a pulsating variable star with the Baade-Wesselink method \citep{1926AN....228..359B, 1946BAN....10...91W}. 
Although the Gaia DR2 has released the distances of approximately 1.3 billion sources \citep{2018A&A...616A...1G}, the Baade-Wesselink method can independently determine the distance of pulsating RRL stars in an accurate way \citep{2005AJ....129.2257M}.
The two different methods can mutually be checked with each other. 

With the wealthy photometric data delivered by TESS and {\sl Kepler}, together with the ground-based RVs, the $p$-factor can be derived for a large amount of RRL stars.
Their values can be applied to all RRL stars in a statistic way, in particular RRL stars residing in globular clusters, to verify the period-luminosity relationship of RRL stars.

\section{Summary}
\label{sec:sum}

In this work, we report the study of the RRab star EPIC~212182292 based on the {\sl K}2 photometry and LAMOST spectroscopy.
The light curves of EPIC~212182292 over a duration of about 80 days from {\sl K}2 and 55 high-quality medium-resolution spectra from LAMOST are collected.

A precise fundamental frequency is resolved as $f$ = 1.8323(17) d$^{-1}$, which corresponds to the period of $P$ = 0.5458(5) days, with the amplitudes of the light curves and RV curves of 0.2968 mag and 23.11\,km s$^{-1}$, respectively. 
The peak-to-peak variations $A$ are 0.8455 mag and 60.83\,km s$^{-1}$ in brightness and RV, respectively.
The rising times, $RT$, are crucial to the shape of RRL stars, whose values are 0.1700 and 0.16 rad for the light curve and RV curves, respectively.

From Fourier transforms, we derive the amplitude ratios of $R_{21} = 0.5115(15), 0.490(8)$ and $R_{31} = 0.3249(20), 0.279(7)$, and the phase difference of $\varphi_{21}=2.7550(20), 2.764(16)$ and 
$\varphi_{31}=5.7194(25), 5.719(31)$ for the light curves and RV curves, respectively (See Table\,\ref{tab:fourier}). 
We also provide the atmospheric parameters of EPIC~212182292 as \teff\ = 6569$\pm$200\,K, \logg\ = 2.46$\pm$0.05\,dex, and [M/H] = -0.40$\pm$0.55\,dex from the LAMOST spectra.

Based on the estimated parameters in Section \ref{subsec:pae}, we construct a series of time-dependent convection models with MESA-RSP. 
A survey grid consisting of 154752 models is built to search for the optimal models of EPIC~212182292. 
Constrained by the period of the fundamental with $\Delta P \leq 0.0005$\,d, the parameter space is significantly reduced to 356 models. 
Then, we take the convective parameters $\alpha_s$ as a variant of four sets into account to mimic the precise shape of the observational curves.

Four optimal models have been chosen based on the minimum values of the merit function of $d$ and $\sigma$.
Comparing the differences among the 4 selected models, we conclude that the radiative cooling parameter $\gamma_r=0$ should be assigned for this RRL stars.
The most optimal model M4 constructs well fitted light curves and RV curves, which shows similar profiles of the observational ones (Figure \ref{fig:lcv} ). 
The projection factor $p$ is estimated to be $1.11\pm0.02$, which could be referred to the investigations of other RRL stars. 
The fundamental parameters which we determine are $M = 0.48\pm0.03$\,M$_{\odot}$, $L=42\pm2$\,L$_{\odot}$, \teff $=6846\pm50$\,K, \logg\ = $2.79\pm0.01$\,dex, and $Z = 0.006\pm0.002$.
Those values are smaller than those derived from the evolutionary models for ZAHB stars, which indicates that
significant mass loss might not be treated correctly by the evolutionary models, for instance, mass taken away caused by radial motion and shock wave at the surface during pulsating cycles.

As phase II of the LAMOST-\kepler\ project is collecting more spectra for high-amplitude variable stars not merely RRL, the seismic modelling of such variables can better constrain the knowledge of our understandings to the hydrodynamic processes in pulsating stars in particular with large amplitudes. In a long way, as opened by large surveys of photometry and spectroscopy, a statistic view on the model determined parameters will help one to imporve the period luminosity relationship, or even draw clues for the Blazhko effect.

\section*{Acknowledgements}
We thank the anonymous referee and J. Nemec for useful comments and suggestions to improve this manuscript. We acknowledge support from the National Natural Science Foundation of China (NSFC) through grants 11833002, 12090040, 12090042 and 11903005. WZ is supported by the Fundamental Re search Funds for the Central Universities. 
The Guoshoujing Telescope (the Large Sky Area Multi-object Fiber Spectroscopic Telescope LAMOST) is a National Major Scientific Project built by the Chinese Academy of Sciences. Funding for the project has been provided by the National Development and Reform Commission. 
LAMOST is operated and managed by the National Astronomical Observatories, Chinese Academy of Sciences. 
JXW and BZ hold the LAMOST fellowship as a Youth Researcher which is supported by the Special Funding for Advanced Users, budgeted and administrated by the Center for Astronomical Mega-Science, Chinese Academy of Sciences (CAMS).

\section*{Data Availability}
The data underlying this article will be shared on reasonable request to the corresponding author.
 



\bibliographystyle{mnras}
\bibliography{epic212182292-1.4} 








\bsp	
\label{lastpage}
\end{document}